%% file: SFM_v241205.tex
\documentclass[hidelinks,12pt]{article}
\usepackage[utf8]{inputenc}
\usepackage[english]{babel}
\usepackage[a4paper,left=1.9 cm, right= 1.9 cm, top=2.3 cm, bottom=2.3 cm]{geometry}
\usepackage{color}
\usepackage{amssymb,amsmath, amsfonts}
\usepackage{latexsym}
\usepackage{graphicx}
\usepackage{rotating} 
\usepackage[onehalfspacing]{setspace}
\usepackage{booktabs}  
\usepackage{caption} 
\usepackage{subcaption}
\usepackage{natbib}
\usepackage{threeparttable}

\usepackage[scaled]{helvet} 
\usepackage{courier} 
\normalfont
\usepackage[T1]{fontenc}

\usepackage[dvipsnames, table]{xcolor} 
\definecolor{DarkerPineGreen}{RGB}{0, 90, 80} 
\definecolor{dukeblue}{rgb}{0.0, 0.0, 0.61}
\definecolor{darkblue}{RGB}{10, 10, 100}
 \definecolor{siena}{rgb}{0.91,0.45,0.32} 
\definecolor{darkred}{rgb}{0.8,0,0}
\definecolor{darkpowderblue}{rgb}{0.0, 0.05, 0.5}
 \definecolor{dpd2}{rgb}{0.0, 0.043, 0.43}
 \definecolor{siena}{rgb}{0.91,0.45,0.32}
\definecolor{darkblue2}{HTML}{1e3986}
\definecolor{darkred2}{HTML}{B22222}
\definecolor{darkgreen2}{HTML}{1B9E77}

\usepackage[pdftex,bookmarks,colorlinks]{hyperref}
\hypersetup{
  colorlinks,
  citecolor=RoyalBlue,
  linkcolor=ForestGreen,
  urlcolor=NavyBlue}

\begin{document}
	\title{\LARGE \textbf{\color{dpd2} Unveiling True Talent: \\ The Soccer Factor Model for Skill Evaluation} \bigskip}

	\author{Alexandre Andorra\thanks{%
Corresponding author,  \href{mailto:alex.andorra@pymc-labs.com}{\texttt{alex.andorra@pymc-labs.com}}.} \\ Miami Marlins, PyMC Labs \and \hspace*{-0.3cm} Maximilian G\"obel
	\\ Brain} 

			\date{
			\bigskip
			\bigskip
			\bigskip
			\small
			First Draft: April 16, 2024 \\
			This Draft: \today \\ 
			}

		\maketitle

		 \begin{abstract}
\noindent Evaluating a soccer player's performance can be challenging due to the high costs and small margins involved in recruitment decisions. Raw observational statistics further complicate an accurate individual skill assessment as they do not abstract from the potentially confounding factor of team strength. We introduce the Soccer Factor Model (SFM), which corrects this bias by isolating a player’s true skill from the team’s influence. 
We compile a novel data set, web-scraped from publicly available data sources. Our empirical application draws on information of 144 players, playing a total of over 33,000  matches, in seasons 2000/01 through 2023/24.
Not only does the SFM allow for a structural interpretation of a player's skill, but also stands out against more reduced-form benchmarks in terms of forecast accuracy.
Moreover, we propose Skill- and Performance Above Replacement as metrics for fair cross-player comparisons. These, for example, allow us to settle the discussion about the GOAT of soccer in the first quarter of the twenty-first century.

		\end{abstract}		 
		 
		\thispagestyle{empty}

\bigskip

		\noindent {\bf Acknowledgments}:   For helpful comments we thank, without implicating, Chris Fonnesbeck, Ravi Ramineni, Osvaldo Martin, Luciano Paz, Aaron MacNeil, Patrick Ward, Paul Sabin and Luke Bornn.

\bigskip


\smallskip

\normalsize


	\clearpage

	\setcounter{page}{1}

\section{Introduction} \label{sec:intro}

The signing of a new player is a delicate decision that can impact a team far beyond its sole performance on the pitch. 
The ability for a  manager or scouting team to discriminate between a nugget and a lemon is hence crucial. 
Historically, and especially in European football -- that game that the Americans call ``soccer'' -- this process was rather discretionary in nature. But similar to the world of investing, one could also follow a more systematic and model-driven approach. In both cases, whether discretionary or systematic, the raw ``stats" of a player are obviously a first point of reference. But those raw statistics, i.e. a player's observed performance, can be misleading as they might be confounded by the influence of the team. Said differently, the \textit{observed performance} is a convolution of a player's skill and the team's effort or strength.
Using thus the observed performance as the decisive metric can lead to a bias in the perception of the actual strength of the player, culminating in spurious conclusion and fatal misjudgement. This bias can be either upward, which can turn out to have negative consequences for the team, or downward, which causes a player to being not as much appreciated as he actually deserves. Hence, knowledge of the bias can prevent teams from overpaying for individual players, or finding a hidden gem.
It is thus tantamount for a manager to have a procedure at hand that can discriminate between the contribution to a player’s observed performance that is due to the strength of the team, and the contribution that comes from the player himself, i.e. his own skill.
Said differently, having a procedure at hand that can answer the question whether it is the player who is carrying the team, or rather the team who is carrying the player, can give a team a competitive advantage.

\vskip 5pt
\noindent{\textsc{\textbf{The Soccer Factor Model.}}}  
To this end, we propose the  Soccer-Factor-Model (SFM) -- a model that allows the manager to strip off the team's strength from the player's observed performance in order to extract a player's unconfounded skill -- the player's $\alpha$.

\vskip 5pt
\noindent{\textsc{\textbf{Relation to Literature on Financial Asset Pricing.}}}  
The SFM itself is inspired by the financial academic literature on asset-pricing, in particular, by the literature on factor investing that followed the seminal work of \cite{FF1992,FF1993}.
The underlying concept rests on the idea that certain ``factors'' can explain the cross-section of expected stock returns. Hence, there are systematic -- common to all stocks -- factors that help explain the movements in observed returns. 

Based on this core concept, this strand of literature has branched off into evaluating the skill of an investment manager \citep{CogginFabozziRahman1993,FamaFrench2010,BerkBinsbergen2015}. ``Skill'' in this context relates to the question whether the performance of a portfolio manager  is due to his ability to pick stocks (or assets in general) in an intelligent manner, i.e. buying future winners and shorting future losers, or whether the observed performance could just be replicated by a (linear) combination of prominent \textit{factors}. In a simple linear regression framework -- that is deployed in that kind of literature -- which regresses a manager's historical return series on prominent factors, the \textit{skill} component is nothing else than the intercept, and henceforth also called the manager's $\alpha$.

Similar to the literature on evaluating investment managers' skill, the SFM makes use of ``factors'' to proxy for a team's strength and elicit a player's $\alpha$. More specifically, factors are constructed either as \textit{absolute} variables, such as the team's rank in the number of goals scored, or as \textit{relative} variables, such as the difference in points between the player's team and the opposing team that the player is facing in the upcoming match.

The whole idea of the SFM is thus to condition a player's observed performance on the strength of the team that he is playing for. This allows any data science team to split a player's observed performance into a player's individual skill and team effort.

\vskip 5pt
\noindent{\textsc{\textbf{SFM: Contributions.}}}  
But the SFM is not confined to the sport of soccer only. It is adaptable to any type of (team) sport, and its output is multifaceted.
More specifically, the contribution of the SFM is threefold: it provides the modeler with (i) a prediction for how a player is expected to perform in a given match , (ii) a quantification of how much a player's skill -- his $\alpha$ -- and the individual constituents thereof, have contributed to his observed performance, and (iii) the possibility to compare players to one another or to whole groups of players, after having generated a level-playing field that accounts for any confounding factors such as differences is strength across teams.

\vskip 5pt
\noindent{\textsc{\textbf{Relation to Existing Literature on Sports Analytics.}}}  
The latter directly speaks to one of the hallmarks of the SFM, namely to rank players cross-sectionally or compare them against a certain benchmark. In the sports analytics literature one such related concept is known by the acronym \textit{WAR}, short for \textit{wins above replacement}. WAR is based on the idea of comparing a player's of interest (let's call him an \textit{elite player} (EP)) contribution to winning a game against the contribution to winning a game that a \textit{replacement level player} (RLP) would make. In general, an RLP could be any type of \textit{real} player that the modeler deems a relevant benchmark or a composite of several players\footnote{For example in Finance, such a composite would be the S\&P 500, which serves as the benchmark for the majority of academic research and portfolio managers in practice.}. One of the early papers in the sports analytics literature on WAR, defines the RLP as ``\textit{the typical player that is readily accessible in the absence of the player being evaluated.}'' \citep{BaumerJensenMatthews2015}. 
Although, the framework of WAR had already circulated on various online blogs\footnote{See  \cite{BaumerJensenMatthews2015} for references.}, in their application to baseball (MLB) \cite{BaumerJensenMatthews2015} were the first to attach a real statistical component, and most importantly by providing a quantification of \textit{uncertainty} for their WAR estimate. Different from the SFM, however, which derives its uncertainty estimates from the Bayesian framework, \cite{BaumerJensenMatthews2015} resorted to resampling methods, which is a popular method for uncertainty quantification in frequentist settings of all kinds of applications (see e.g \cite{DieboldEtAl2023}) -- though kindly dubbed as ``a ''poor man's'' Bayes posterior'' by \cite{HastieTibshiraniFriedman2017}.

\cite{YurkoVenturaHorowitz2019} then adapted the concept of WAR for American Football and introduced \textit{nflWAR}. 
The general idea of \textit{openWAR} and \textit{nflWAR} is very similar, and complies with the concept of WAR: first, calculate each player's contribution to winning a game, and then compare these contribution to some benchmark, the RLP. Yet, it is the modeling of the first step, where these studies differ, as the proposed models are tailored to the specific sport. 

\vskip 5pt
\noindent{\textsc{\textbf{Skill-/Performance Above Replacement.}}}  
In the spirit of these studies and the concept of WAR in general, we propose a fairly similar metric that distills the output of the SFM into easily accessible and interpretable measures of outperformance. In particular, in Section \ref{sec:sar_par} we introduce SAR (\textit{skill above replacement}) and PAR (\textit{performance above replacement}). Different from WAR though, these metrics do not measure a player's contribution to \textit{winning} a game, but rather how his \textit{skill}, SAR, or general \textit{performance}, PAR, compares to a given benchmark. 

The rest of the paper is structured as follows: in Section \ref{sec:sfm}, we introduce the SFM and discuss the relevant modeling choices. In Section \ref{sec:sar_par} we go one step further by introducing the above mentioned SAR and PAR -- metrics that bundle the output of the SFM to compare the skill and performance across players. We present the empirical results in Section \ref{sec:emp} before concluding in Section \ref{sec:concl}.

\section{SFM: The Model} \label{sec:sfm}

When judging a player based on his performance, the obvious first choice is to look at his ``stats'' -- his \textit{observed performance} (OP). However, what we measure with OP is actually a convolution of a player's \textit{skill} ($\alpha$) and \textit{team performance} (TP). Hence, to really understand, which part of OP is actually attributable to a player's own performance, one has to disentangle $\alpha$ from TP. To do so, we model OP as a function of $\alpha$ and TP, such that:

\begin{align} \label{equ:op_alpha_tp}
	\text{OP} = g\left(\alpha,\text{TP}; \theta\right) \; .
\end{align}

where $\theta$ is a vector of parameters to be estimated.
Without loss of generality, we assume that $g\left(\cdot\right)$ is linear in its arguments.  

In general, there are two schools of thought that propose different frameworks for how to go about modeling OP: the frequentist and a Bayesian way of thinking. We opt for the Bayesian framework, which, in contrast to the frequentist approach, does not treat the observed data as a random variable, but assume the coefficients of $g\left(\cdot; \theta\right)$ that best describe the observed data, to be a draw from some underlying ground-truth distribution. This framework allows us to not only easily quantify the degree of uncertainty of our point estimates, but also to incorporate some domain knowledge into our model, in particular by imposing restrictions on the parameters $\theta$.

Under the Bayesian framework, estimating Equation \eqref{equ:op_alpha_tp} boils down to the following:

\begin{align} \label{equ:op_bayesian}
	P\left(\theta \; | \; \text{OP},\alpha, \text{TP}\right) \propto P\left(\text{OP} \; | \; g\left(\alpha,\text{TP}; \theta\right) \right) \times P\left(\theta\right) \quad ,
\end{align}

where $\theta$ is the collection of parameters that describe the function $g\left(\cdot; \theta\right)$. $P\left(\theta\right)$ is the so called ``\textit{prior}'', and describes the distribution of our parameters that we think encompasses a reasonable set of values that $\theta$ can take on, such that the function $g\left(\cdot; \theta\right)$ is an accurate description of our target variable OP. For a given set of values $\hat{\theta}$, the likelihood function $P\left(\text{OP} \; | \; \hat{\theta}, \alpha, \text{TP} \right)$ then gives us an indication of how reasonable our proposed values $\hat{\theta}$ are for $g\left(\alpha,\text{TP}; \theta\right)$ to describe the observed data -- here: OP.

Our object of interest is $P\left(\theta \; | \; \text{OP}, \alpha,\text{TP}\right)$: the posterior distribution of $\theta$. This describes the distribution of the parameters of function $g\left(\cdot; \theta \right)$ that emerges from having updated our prior beliefs about the distribution of $\theta$, $P\left(\theta\right)$, with the likelihood of that prior distribution being an accurate description of how our observed data is actually distributed ($P\left(\text{OP} \; | \; \theta, \alpha, \text{TP} \right)$).

In order to estimate $P\left(\theta \; | \; \text{OP}, \alpha,\text{TP}\right)$ in Equation \eqref{equ:op_bayesian}, we have to determine the following components: (i) the distribution that is an accurate representation of OP -- which in Equation \eqref{equ:op_bayesian} determines the shape of the likelihood function $P\left(\text{OP} \; | \; \hat{\theta}, \, \alpha, \text{TP} \right)$, (ii) the data-generating-process (DGP) of the two components, $\alpha$ and TP, of function  $g\left(\cdot; \theta \right)$, and lastly (iii) the prior distribution(s) for all parameters in $\theta$, which in our model is the union of parameters required to model $\alpha$ and TP. 

The following subsections describe the choices we made for each of these three components of the SFM.

\subsection{Determining the Likelihood} \label{sec:likelihood}
As mentioned earlier, the SFM is an agnostic framework that is not confined to a particular application. Thus, OP is not restricted to a specific performance measure, but can be easily adapted to deal with performance measures of various player types. 

To showcase the general workings of the SFM, we opt for assessing the skill of strikers and set our target variable $OP_{i,m,s}$ to be the number of goals scored by player $i$ in a given match $m$ of a particular season $s$.\footnote{Alternatively, one could also set $OP_{i,m,s}$ to be a binary variable taking on the value of 1 if player $i$ had scored a goal in game $m$ of season $s$, and 0 otherwise. Thus, one would use the SFM to model the probability of player $i$ scoring a goal in a given match. Yet, this binary classification exercise lumps a lot of information that the number of goals conveys and cushions an enormous amount of heterogeneity among players that would also squish the heterogeneity across our $\mathbf{\alpha}$. }

Our choice for $OP_{i,m,s}$ being the number of goals in a given match, necessitates several considerations regarding the choice of the likelihood function. First of all, $OP_{i,m,s} \in \mathbb{Z}^{+}_0$, which already eliminates a whole set of candidate distributions. A standard choice to satisfy the support of $OP_{i,m,s}$ is to model $P\left(\text{OP} \; | \; g\left(\alpha,\text{TP}; \theta\right) \right)$ as a Poisson distribution. However, in our sample, in only 1.5\% of cases do we observe our candidate players to have scored 3 goals or more in a single game. Thus, our $OP_{i,m,s}$ calls for a distribution with a very flat right tail, rendering a Poisson distribution not appropriate.

Given that occurrences of 3+ goals are rather rare, we set all observations $OP_{i,m,s} > 3$ equal to 3, such that:

\begin{align*}
	OP_{i,m,s}=
	\begin{cases}
			3, & \text{if} \quad OP_{i,m,s} \geq 3 \\
			OP_{i,m,s}, & \text{otherwise} \quad.
	\end{cases}
\end{align*}

This procedure is in essence \textit{winsorization}, which is for example often used by economists working with firm-level data to perform \textit{outlier} treatment, and performs a truncation of the tails of the distribution (here at the 98.5$^\text{th}$ percentile) without throwing away all the information that is hidden in the corresponding observations.

Once done, we are left with four different categories, such that $OP_{i,m,s} = \left[\,0,1,2,3+\, \right] $. With an upper bound on $OP_{i,m,s}$ and a meaningful order of categories, we set our target variable to follow an ordered logistic distribution 

\begin{align} \label{equ:likelihood}
	OP \sim \text{OrderedLogistic}\left(c \; | \; g\left(\cdot;\theta\right), \zeta\right) \quad ,
\end{align}

where $c \in \left\lbrace1,...,C\right\rbrace$ is the number of categories, $g\left(\cdot;\theta\right)$ is our prediction for $OP_{i,m,s}$, and $\zeta \in \mathbb{R}^{C-1}$ is a set of cutoff points to split the values of $OP_{i,m,s}$ into each of the $C$ categories.

\subsection{Determining the DGP for Player-Skill and Team-Effort} \label{sec:alpha_TP}

In this section, we describe the DGP that governs our predictive function $g\left(\alpha,\text{TP}; \theta\right)$, which is composed of two components, $\alpha$ and TP, and parameterized by $\theta$.

To recall, $\alpha$ describes the individual part of a player's observed performance, i.e. his skill, and is our key object of interest. In contrast, TP captures that part of the observed performance, that has to be ascribed to the (relative) strength of the team that the player is playing for. This latter component is composed of \textit{factors} that capture \textit{systematic} differences between the player's team and the opponent, influencing the player's OP, but not his innate skill. 

\subsubsection{Team-Performance: The Factor Structure} \label{sec:factors}
Even though we are mostly interested in a player's $\alpha$, TP and its very own DGP play an essential role in the SFM. The plain OP of a player is in general biased due to the natural heterogeneity in strength across teams. Thus, our \textit{factors} are intended to account for this bias by capturing exactly these cross-sectional, i.e. cross-team, differences. Formally, we define TP as follows:

\begin{align} \label{equ:factors}
	TP_{i,m,s} \equiv \beta_1 X_{1,i,m,s} \; + \; \beta_2 X_{2,i,m,s} \; + \; ... \; +\; \beta_F X_{F,i,m,s} \; ,
\end{align}

where $TP_{i,m,s}$ is the team performance of player $i$'s team being modeled as a linear combination of $f = 1, ..., F$ factors that capture the strength of player $i$'s team \textit{relative} to the adversary team that is faced in match $m$. One can think of these factors as $F$ categories which the modeler deems sufficient to span (or capture) the differences between two teams. $\beta_f$ is the corresponding loading or weight on factor $f$ and is one member of the parameter vector $\theta$ in $g\left(\cdot; \theta\right)$ that we will estimate via Equation \eqref{equ:op_bayesian}.

For simplicity, let us denote $\mathbf{X}_{i,m,s}$ as the vector of all factors $F$ associated with player $i$ for a particular match $m$ in season $s$.
Note again that $\mathbf{X}_{i,m,s}$ is measured in real-time. That is, our features only incorporate information between the start of season $s$ and the real-time kick-off of match $m$ in the very same season $s$. For example, there might be special cases in which match $m^{*}$ was supposed to be played on game day $G$, but had to be postponed, with the actual kick-off ($G^{*}$) now being set in between the official game days $G+3$ and $G+4$. Thus, for all gamedays until and including $G+3$, the constituents of $\mathbf{X}_{i,m,s}$ will \textit{not} incorporate information of $m^{*}$, as it's actually kick-off is set to a date \textit{after} $G+3$. However, for $m^{*}$ itself, we will account for all information that had become available up until the actual kick-off of match $m^{*}$, hence information of gamedays up until and including $G+3$.

\subsubsection{Player-Skill: More than just $\alpha$}\label{sec:alpha}

If we were to follow the stream of asset-pricing literature that the SFM builds upon, we would just model a player's skill as a single component: $\alpha$. 
In that case, Equation \eqref{equ:op_alpha_tp} would be specified as:

\begin{align*}
	\text{OP}_{i,s,t} = g\left(k=4 \; | \; \eta = \alpha_{i,m,s} \; + \; \beta_1 X_{1,i,m,s} \; + \; \beta_2 X_{2,i,m,s} \; + \; ... \; +\; \beta_F X_{F,i,m,s} \; ,  c \right) \; ,
\end{align*}

where $g\left(\cdot\right)$ is now defined as the parameterized ordered logistic distribution as described in Equation \eqref{equ:likelihood}. Here, $\alpha_{i,m,s}$ would describe the skill of player $i$ -- nothing more. 

However, we want to allow for a player's skill to be a much richer assemblage of several components. A player's skill might be subject to some kind of evolution across several seasons -- or not. By this, we refer to the idea of ``grit'' , whereby some talented players may rise up the ranks very fast and very early on in their career, but then deteriorate sharply as their career progresses. Others in contrast, not being endowed with as much talent, may progress steadily over the years, peaking later on in their careers. We dub this the ``maturity effect''.
Another component of skill is the extent to which a player is subject to within-season fluctuations, e.g. performing well early on in the season but then running out of steam as the season progresses.

Incorporating all these considerations requires to represent a player's skill as much more than just a single player-specific intercept. Thus, we model $\alpha_{i,m,s}$ as follows:

\begin{gather} \label{equ:alpha}
	\begin{split}
	\alpha_{i,m,s} \equiv \underbrace{\mu_b \; + \; \delta_{b,i}}_{\text{player effect}} \; + \;  \underbrace{\underbrace{h^W\left( D_{i,0:m,s} \right)}_{\text{within-season effect}} \; + \;  \underbrace{h^C\left( N^S_{i,s} \right)}_{\text{cross-seasonal effect}}}_{\text{maturity effect}} \quad	
	s.t. \quad \sum^I_i \delta_{b,i} = 0 \; , \\ \\
	\mu_b \sim N\left(0; \sigma^2_{\mu_b}\right) \qquad \text{with} \qquad \sigma_{\mu_b} = \sqrt{\iota_{\sigma_b}^2 \, + \, \frac{\sigma_{\sigma_{\mu_b}}^2}{N_P}}  \qquad \text{and} \qquad \sigma_{\sigma_{\mu_b}} \sim \text{Exp}\left(1\right)
	\end{split}
\end{gather}

where $\mu_b$ captures the baseline skill level across all players ($N_P$). Some first-order differentiation is then introduced by $\delta_{b,i}$, which measures the diversity in the baseline skill across players. In essence, the first two terms, the ``player effect'', capture a player's \textit{natural talent}. To discipline the model, the $\delta_{b,i}$ are constrained to sum to zero across all players, having the effect that the player with average talent will be assigned $\delta_{b,i} = 0$. $\iota_{\sigma_b} = 5$ is a constant and serves as a variance shifter for $\mu_b$.

 The ``maturity effect'', a measure of \textit{grit}, is modeled by the second and third components, which describe within- and cross-seasonal fluctuations respectively.
$D_{i,0:m,s}$ is the number of days between the first game that player $i$ has played and the current match $m$ in a given season $s$, and $N^S_{i,s}$ counts the number of total seasons player $i$ has been observed in our sample up until and including season $s$. The functions $ h^W\left(\cdot\right)$ and $ h^C\left(\cdot\right)$ are Hilbert-Space decompositions of a Gaussian Process (HSGP).
Now, this last modeling choice may raise questions that deserve explanation and justification: (i) why an HSGP and not a simple Gaussian Process (GP)?, and (ii) why a GP in the first place?

To build intuition for our choices, we answer these questions, in reverse order: first, any career path is hardly ever a straight line. There are short-term fluctuations, which we capture by within-season effects, and more longer-term trends or cycles, which we capture by cross-seasonal effects. For example, within a season, a player's performance may be subject to sudden changes, stemming from physical or psychological issues, while a players' overall career might be highly multi-modal. Clearly, these fluctuations are of non-linear nature that are hard to be modeled explicitly. Similar to machine- and deep-learning algorithms, GPs offer a convenient way out of this dilemma, as they belong to the class of \textit{universal function approximators} \citep{MicchelliXuZhang2006}: via a combination of several basis-functions, the observed data can be modeled in a flexible way without the need to embark on excessive feature-engineering, generating \textit{any} higher-order transformations (polynomials) of our features $D_{i,0:m,s}$ and $N^S_{i,s}$, to capture potential nonlinearities. In fact, via kernel functions, GPs represent a potentially \textit{infinite} dimensional feature vector $\phi^W\left(D_{i,0:m,s}\right)$ and $\phi^C\left(N^S_{i,s}\right)$ of transformations of our predictor variables in a \textit{finite} space \citep{HastieTibshiraniFriedman2017}. This kernel representation effectively approximates each observation of our target $OP_{i,m,s}$ by a function $h^W\left(\cdot\right)$ and $h^C\left(\cdot\right)$ over the corresponding input (or feature) spaces.

Thus, if we were to approximate the components of the maturity effect via polynomials, e.g.:

\begin{align*}
	\phi^W_{i,m,s} = \left(1, D_{i,0:m,s}, D^2_{i,0:m,s}, D^3_{i,0:m,s} \right) \; , \\
	\phi^C_{i,s} = \left(1, N^S_{i,s}, \left(N^S_{i,s}\right)^2, \left(N^S_{i,s}\right)^3 \right) \; ,
\end{align*}

we would need to set a prior on each element in $\phi^W_{i,m,s}$ and $\phi^C_{i,s}$, such that for example $\phi^W_{i,m,s} \sim MVN\left(\mathbf{\mu_4},\mathbf{\Sigma_{4,4}}\right)$. In this probably more familiar case, we would sample a four-dimensional vector of weights from a four-dimensional multivariate normal distribution. Modeling the within- and cross-seasonal effects via GPs however, will not result in sampling scalar values for the loadings on each \textit{feature}, but rather in sampling for each \textit{observation} (here: match $m$) a distribution defined by a mean and covariance \textit{function}. Taking the within-season effect as an example, the mean would be described by $\mu^W\left( D_{i,0:m,s} \right)$, and the covariance function by a kernel, $ k^W \left( D_{i,0:m,s},D_{i,0:m^{'},s} \right)$, where $m^{'}$ defines any other match that player $i$ has played in season $s$ but $m$. Thus, the prediction for the within-season effect of player $i$ in match $m$ of season $s$, is generated by a function $h\left( D_{i,0:m,s} \right)$ that is drawn from a (normal) Gaussian distribution with mean $\mu\left( D_{i,0:m,s} \right)$ and by a vector of covariances $ k\left( D_{i,0:m,s},D_{i,0:m^{'},s} \right)$ that describe the similarity between the input features (here only $D_{i,0:m,s}$) of match $m$ and match $m^{'}$, where $m^{'} = \left\lbrace 1, ..., M \, \diagdown \, m \right\rbrace$:

\begin{gather*}	
			h^W \sim \mathcal{GP}\left(\mu^W\left( D_{i,0:m,s}  \right), \iota^2 \; k^W \left( D_{i,0:m,s} ,D_{i,0:m^{'},s}  \, | \, \mathit{l}\right) \right) \; , \\
			\mu^W\left( D_{i,0:m,s}  \right) = \mathtt{E}\left[h^W\left(D_{i,0:m,s} \right)\right]\; , \\
			k^W \left( D_{i,0:m,s} ,D_{i,0:m^{'},s}  \right) = \mathtt{E}\left[\left(h^W\left(D_{i,0:m,s} \right) - \mu^W\left(D_{i,0:m,s}  \right)\right) \left(h^W\left(D_{i,0:m^{'},s} \right) - \mu^W\left( D_{i,0:m^{'},s}  \right)\right)\right]\; .	
\end{gather*}

Extending this to each of the games $m = 1,...,M$ played by player $i$, $\mathbf{h}^W\left(D_s\right)$ follows a multivariate Gaussian distribution $h^W \sim MVN\left(\mathbf{\mu}^W_s, \mathbf{K}^W_s\right)$ where  $\mathbf{\mu}^W_s$ is the vector of means for each game $m$ in season $s$, and $\mathbf{K}^W_s$ is the corresponding $M \times M$ covariance matrix, with $K^W_{i,j;s}$  being the covariance between games $m_i$ and $m_j$ in season $s$.

Beyond being flexible estimators, GPs come with a great degree of interpretability. The \textit{lengthscale} parameter ($\mathit{l}$), which is an integral part of the kernel, i.e. the covariance function, determines the degree of memory or time-dependence that a particular observation draws on -- in a time-series context also often referred to the degree of autocorrelation in the process \citep{DixonHalperinBilokon2020}. It also determines how smooth (larger values) or wriggly (lower values) the estimated function $\hat{h}^W\left(\cdot\right)$ will be \citep{Murphy2023}. The \textit{amplitude} ($\iota$) sets the range within which our prediction ($\widehat{OP}_{i,m,s}$) for the current observed target value ($OP_{i,m,s}$) may vary, thus pinning-down the expressivity of the model. Taken together, these two parameters define the area ($A$), which the GP will consider to make a prediction $\hat{h}^W\left(\cdot\right)$ that ``fits'' all the observations $OP_{i,m \in A,s}$ most accurately.\footnote{In essence, this is like fitting local linear regression models.}

The drawback of plain GPs is of computational nature and rooted in the necessity of having to invert a potentially high-dimensional covariance matrix of dimension $N \times N$, where $N$ denotes the number of observations \footnote{Though, as stated in this excellent post by \href{https://juanitorduz.github.io/hsgp_intro/}{Juan Orduz}, in practical applications one would opt for the Cholesky decomposition of $\mathbf{K}_s$ instead of trying to invert it. Still an $\mathcal{O}\left(n^3\right)$ problem. }. This is further aggravated by the need to perform this inversion in each sampling step \citep{RiutortMayolEtAl2022}. 
HSGPs provide a solution to this problem by representing the kernel by the sum of its eigenvalues and eigenvectors.\footnote{Again very nicely laid out by \href{https://juanitorduz.github.io/hsgp_intro/}{Juan Orduz}.}
With our covariance matrix $\mathbf{K}_s$ being symmetric and semi-positive definite, choosing a kernel that is stationary, such as the Mat\'ern class, allows us to represent the covariance function via its spectral densities. Thus, the HSGP approximation of a kernel $k\left(\cdot,\cdot\right)$ of the class of Mat\'ern covariance functions for the within-season effect can be written as:

\begin{align} \label{equ:hsgp_kernel}
	k^W \left( D_{i,0:m,s},D_{i,0:m^{'},s} \right) = \sum^B_{b=1} \mathfrak{s}_\chi\left(\sqrt{\lambda_b}\right) \; \phi_b\left(D_{i,0:m,s}\right) \; \phi_b\left(D_{i,0:m^{'},s}\right) \; ,
\end{align}

where $\mathfrak{s}_\chi\left(\cdot\right)$ is the spectral density of the covariance function with corresponding hyperparameters $\chi$. The eigenvalues and eigenvectors of the aforementioned Laplacian operator are denoted by $\lambda_b$ and $\phi_b$. 
As stated earlier, HSGPs are only an approximation of GPs, which is why we set the number of basis functions to a finite number $B < \infty$.

The Mat\'ern kernel is composed of two hyperparameters $\chi = \left\lbrace\nu, \mathit{l} \right\rbrace$, where $\mathit{l}$ is the aforementioned lengthscale parameter, and $\nu$ denotes the order of the kernel, interacting with $\mathit{l}$ to pin down the roughness of kernel function. One can think of $\nu$ as the number of polynomials of $\lambda_b$, approximating the spectral density $\mathfrak{s}_\chi\left(\cdot\right)$. Following the recommendation in \cite{RasmussenWilliams2006} we set $\nu = 5/2$, and as our input space is just one-dimensional, the spectral density of our Mat\'ern kernel can be written as:

\begin{align} \label{equ:hsgp_kernel_spectral}
	\mathfrak{s}_\chi\left(\sqrt{\lambda} \, | \,  \mathit{l} \right) = \frac{2\sqrt{\pi} \; \Gamma(3) \;5^{\frac{5}{2}}}{\frac{3}{4}\sqrt{\pi} \; \mathit{l}^5} \left(\frac{5}{\mathit{l}^2} + \lambda^{\mathsf{T}}\lambda \right)^{-3} \; ,
\end{align}

The missing ingredient to compute the spectral density of the kernel is the lengthscale $\mathit{l}$. As mentioned before, it indicates the degree of autocorrelation or memory hidden in the evolution of our target variable OP.
So far, we have specified the \textit{maturity effect} of a player's $\alpha$, as being the sum of a \textit{within-} and \textit{cross-seasonal} component. However, a single \textit{within-season} effect may fall short of differentiating between very short-term fluctuations (e.g. capturing the phenomenon of what in basketball jargon is described as a ``hot hand'', or the tendency to suffer from small injuries) and medium-term fluctuations, such as slow-moving fatigue. Thus, we further distinguish between short- and medium-term fluctuations within the \textit{within-season effect}. Thus, we set three different HSGPs: two for the within-season effect, and one for the cross-seasonal fluctuations. This results in three different lengthscale parameterizations, denoted by $ T = \left\lbrace S,M,L\right\rbrace$, which we set to be described by an inverse Gamma distribution:

\begin{gather*}
	\mathit{l}^T_i = \Gamma^{-1}\left(\alpha^T , \beta^T\right) \quad \text{for} \quad T = \left\lbrace S,M,L\right\rbrace \; , \\
	\alpha^S \,  , \, \beta^S \sim \mathcal{M}\left(\Gamma^{-1}, 2,5 \right) \; ,\\
	\alpha^M \,  , \, \beta^M \sim \mathcal{M}\left(\Gamma^{-1}, 2,5 \right) \, \\
	\alpha^L \,  , \, \beta^L \sim \mathcal{M}\left(\Gamma^{-1}, 15,30 \right) \; ,
\end{gather*} 
where $\mathcal{M}\left(\Gamma^-1, \mathfrak{m}_1, \mathfrak{m}_2 \right)$ indicates that parameters $\alpha$ and $\beta$ are to be drawn from an inverse Gamma with all distributional mass lying in the interval $\mathfrak{m}_1$ and $\mathfrak{m}_2$.

Equipped with an approximation of the kernel, we can finally construct the approximation of our GPs, i.e. our HSGPs. The example for the within-season effect, which is the sum of short- and medium-term fluctuations, is written as:

\begin{align} \label{equ:hsgp}
	h^W\left(D_{i,0:m,s}\right) = \sum^B_{b=1} \left(\sqrt{\mathfrak{s}_\chi\left(\sqrt{\lambda_b} \, | \, \mathit{l}^S_i \right)} \; + \; \sqrt{\mathfrak{s}_\chi\left(\sqrt{\lambda_b} \, | \, \mathit{l}^M_i \right)} \right) \; \phi^W_b\left( D_{i,0:m,s} \right) \; \Delta_b
\end{align}

which is equivalent to saying that the function $h\left(D_{i,0:m,s}\right)$ is described by a zero-mean multivariate normal distribution:  $h^W\left(D_{i,0:m,s}\right) \sim MVN\left(\mathbf{0}, \Phi \mathcal{S}^W \Phi^{\mathsf{T}}\right)$, where $\mathcal{S}^W$ is a $B \times B$ diagonal matrix with $\sum_{t=S,M} \mathfrak{s}_\chi\left(\sqrt{\lambda_b \, | \, \mathit{l}^t_i}\right)$ on the diagonal and $\Phi$ is the $M \times B$ matrix of eigenvectors of the $M$ observations and $B$ basis functions respectively \citep{RiutortMayolEtAl2022}. $\Delta_b$ is the square root of the power spectral density.

\subsection{Setting Priors} \label{sec:priors}

Critics of Bayesian estimation say that setting priors is more of an art than a scientific endeavor, making the model prone to overfitting. Advocates say it allows to incorporate domain knowledge, thus making the model more robust and less prone to overfitting. Regardless of which train of thought one subscribes to, we find the latter well suited for the application of the SFM. Besides all the data-driven checks, which we will refer to at the corresponding instances, domain knowledge clearly serves here as a safeguard against model misspecification: if the SFM puts the GOAT(s) of the respective discipline into the left tail of the skill distribution, something is off, even though the ``true'' model -- whatever that is -- may still not put them within the 99$^\text{th}$ percentile.

To get the Bayesian workflow running, we thus have to take a stance of how we define the priors for each of our parameters that need to be estimated. As laid out in Section \ref{sec:sfm}, we let $\theta$ define the vector that collects all necessary parameters of the model. 
As already mentioned in Section \ref{sec:intro}, a key contribution of this paper is to evaluate a player's skill against the skill of a potential \textit{replacement} candidate, which we called replacement level player (RLP). That is, we take the skill ($\alpha_{i,\cdot,\cdot}$), as given by the SFM, for a given set of elite players (EP) (or players of interest) $i = 1,...,E$ and compare it to the skill of an RLP ($\alpha_{RLP}$). This distinction does however \textit{not} influence the setting of priors.
In what follows, we describe each of the components that ultimately make up this $\theta$ vector\footnote{Here, we omit the subscript $i$, which just served to identify a specific player $i$.}:

Equation \eqref{equ:factors} states that part of the model that describes the team's performance and is a linear combination of $F$ factors, each requiring a loading ($\beta_1, ..., \beta_F$), which will be drawn independently from a normal distribution:

\begin{gather*}
	\beta_f \sim N\left(0, 2.5\right) \quad \text{for} \; f = 1, ..., F \; .
\end{gather*}

In Equation \eqref{equ:alpha}, we need to set priors for six parameters ($\mu_b$, $\delta_b$, $h^W\left(\cdot\right)$, $h^C\left(\cdot\right)$, $\iota^T$, $\mathit{l}^T$), where the superscripts $W$ and $C$ stand for \textit{within-}, respectively \textit{cross-}, seasonal evolution of a player's skill, and the superscript $T = \left\langle S, M \right\rangle$ denotes the short- and medium-term fluctuations of the \textit{within-season effect} and $T=L$ denotes the parameters for the \textit{cross-seasonal effect}:

\begin{gather*}
	\mu_b \sim N\left(0,1\right) \\
	\delta_b \sim \Gamma\left(2,2\right) \\
	\iota^T \sim \text{Exp}\left(-\frac{\text{log}\left(0.01\right)}{2}\right) \\
	\mu_{\mathit{l}} \sim N\left(0,1\right) \\
	\mathit{l}^T= \Gamma^{-1}\left(\alpha^T , \beta^T\right) \\
	\alpha^S \,  , \, \beta^S \sim \mathcal{M}\left(\Gamma^{-1}, 2,5 \right) \\
	\alpha^M \,  , \, \beta^M \sim \mathcal{M}\left(\Gamma^{-1}, 2,5 \right) \\
	\alpha^L \,  , \, \beta^L \sim \mathcal{M}\left(\Gamma^{-1}, 15,30 \right) \\
	h^W \sim \mathcal{HSGP}\left(B=120, c_\mathcal{HSGP}=2.5, \iota^T, \mathit{l}^T   \right) \\
	h^C \sim \mathcal{HSGP}\left(B=120, c_\mathcal{HSGP}=2.5, \iota^T, \mathit{l} ^T  \right) \\
	\Delta_b \sim N\left(0,1\right)
\end{gather*}

This completes the set of parameters that our parameter vector $\theta$ is ultimately composed of. To summarize:

\begin{align*}
	\theta = \left\lbrace \beta_1, ..., \beta_F, \mu_b,  \delta_b, h^W, h^C, \iota^T, \mu_{\mathit{l}}, \mu_{\mathit{l}}, \mathit{l}^T, \alpha^T, \beta^T, \Delta_b \right\rbrace
\end{align*}

Lastly, a parameter that does not belong to the set of coefficients, $\theta$, describing the conditional mean of our target's log-odds ratio, but that plays an integral role in defining our likelihood function in Equation \eqref{equ:likelihood}, is $\zeta$. Our ordered logistic distribution requires the setting of \textit{cut points}, which allow us to assign our model's prediction to a particular outcome category, which in our case is either zero, one, two, or three and more goals. These \textit{cut points}, are described by $\zeta \in \mathtt{R}^{N_P \times N_\zeta}$, where $N_\zeta$ is set to the number of categories minus 1, i.e. $N_\zeta = 3$.

\begin{gather*}
	\zeta_1 \equiv 4  \; , \\ 
	\zeta_2 \sim \zeta_3 \sim 4 + \text{softplus}\left(N\left(\Delta_m , \Delta_s^2\right) \right) \; ,\\
	\Delta_m \sim N\left(\delta_m * 4, 1\right) \; ,\\
	\Delta_s \sim \text{Exp}\left(1\right) \; ,
\end{gather*}
where $\delta_m$ is the difference in the empirical cumulative distribution function of the \textit{ordered} goals scored in our sample.

\section{SFM Add-On: Skill and Performance Above Replacement} \label{sec:sar_par}

As mentioned in the introduction, a popular concept to measure a player's performance or \textit{contribution}, that originated in baseball, is \textit{wins above replacement} (WAR) (see e.g. \cite{BaumerJensenMatthews2015}). 
We adapt this framework for evaluating the performance of football players, and reframe it as \textit{skill} ($SAR$), respectively \textit{performance} ($PAR$), \textit{above replacement}.

These two concepts of $SAR$ and $PAR$ build on the predictions of the SFM and enable the modeler to compare players to one another. $SAR$ differentiates itself from $PAR$ by establishing a level playing field for all players. We can model both these metrics only because the SFM eliminates any potential biases in the observed performance stemming from heterogeneity in strength across teams by separating a player's skill ($\alpha$) from his team's strength (the \textit{factors}).

More specifically, with $PAR$, one will be able to measure how the performance of an EP ($n_e$) compares to an RLP ($n^*_{RLP}$). The $n^*_{RLP}$\footnote{Yes, the $*$ and capitalized subscript are representing the ``hypothetical'' player that each elite player $n_e$ will be benchmarked against.} is constructed simply as the average of all players, $n_{rlp}$, that belong to the set of RLPs: $\mathbb{RLP} = \left\lbrace 1_{rlp},..., N_{rlp} \right\rbrace$. Thus $PAR$ for a given player $n_e$ is defined as:

\begin{align} \label{equ:par}
	\begin{split}
		PAR_{n_e} = \hat{y}_{n_e} - \hat{y}_{n^*_{RLP}} \\
		\hat{y}_{n^*_{RLP}} = \frac{1}{N_{rlp}}\sum_{i \in \mathbb{RLP}} \hat{y}_i \quad ,
	\end{split}
\end{align}

where $\hat{y}_{n_e}$ is SFM's prediction for the observed performance of our elite player of interest $n_e$ and $\hat{y}_i$ is the model's prediction of a given player $i$'s performance who belongs to the set of RLPs, which we denoted above by $\mathbb{RLP}$.

In contrast, $SAR$ corresponds to a comparison of players' skill only. That is, we compare again a given EP, $n_e$, to the hypothetical RLP, $n^*_{RLP}$, but basing our comparison only on $\alpha$, i.e. the skill component. That is, once we have estimated the SFM, which means that we are equipped with the posterior distribution of all our model parameters, we make predictions for all of our players' observed performance ($\hat{y}$) solely based on the skill-component $\alpha$. This is equivalent to establishing a level-playing field, in which we have accounted for team-differentials during estimation to give us an unconfounded estimate of $\alpha$, and ex-post, setting all team-differentials to zero. This allows us to answer two questions: (i) \textit{if all teams were the same, how would a player's performance, based solely on skill, deviate from his observed performance?}, and (ii) \textit{if all teams were the same, how would an elite player compare to the RLP?}.

The difference between $SAR$ and $PAR$ boils down to modelling $SAR$ exactly as $PAR$ in Equation \eqref{equ:par}, with the only exception that $SAR$ bases the prediction for a player's performance -- be it an EP or RLP -- solely on the estimated skill component $\hat{\alpha}$. We denote this prediction as $\hat{y}^\alpha$:

\begin{align} \label{equ:sar}
	\begin{split}
		SAR_{n_e} = \hat{y}^\alpha_{n_e} - \hat{y}^\alpha_{n^*_{RLP}} \\
		\hat{y}^\alpha_{n^*_{RLP}} = \frac{1}{N_{rlp}}\sum_{i \in \mathbb{RLP}} \hat{y}^\alpha_{i,\cdot,\cdot} \quad .
	\end{split}
\end{align}

\vspace{0.5cm}

\section{SFM: Empirical Application}  \label{sec:emp}

In this section, we will showcase the working of the SFM for predicting the number of goals scored per game by a given player. However, it  cannot be stressed enough that the SFM is application agnostic and is easily adaptable to various problem settings with a different target variable.

\subsection{Data}

Our data is sourced from \href{https://www.kicker.de}{\textit{kicker}}, the website of arguably the most popular and reliable German football magazine.\footnote{Similar to the Italian \textit{Gazzetta dello Sport}, the French \textit{L'\'Equipe} or the Spanish \textit{MARCA}.} We compile our sample by web scraping data on each game of the four major European leagues -- Premier League, Bundesliga, Serie-A, and La Liga -- for each of the seasons between 2000/01 and 2023/2024.
This leaves us overall with 34,441 matches, 12,775 individual players from all kinds of positions, and 94,228 goals, out of which we could attribute 96.6\% to the corresponding player. From this universe of observations, we discretionarily select a pool of 144 strikers that we deem to have played an influential part in the game of football throughout the first quarter of the twenty-first century.\footnote{See Table \ref{tab:ListOfPlayers} for a complete list of these players.} This leaves us in total with over 33,000 observations and more than 13,000 goals scored.

To calculate our $SAR$ and $PAR$ metrics post-inference, we split this subset of players into two subgroups, by assigning 42 players to the group of EPs, and 102 players to the group that forms the RLP. Note that the assignment of players into any of the subgroups is entirely subjective, and is based on -- yes -- domain knowledge. To reiterate, the SFM itself is fitted on \textit{all} 144 players. The distinction between EPs and RLPs only enters the SAR and PAR calculations.

Still, it is worth taking a look at the unconditional characteristics of these two subgroups in order to understand the nature of our underlying data. Figure \ref{fig:Goals_byPlayerType_Total} shows the frequency with which a certain number of goals was observed among the two groups of EPs and RLPs. Despite the number of RLPs being more than double the number of EPs, we can already see from this plot, that we have more instances in which EPs score two goals or more than do RLPs. Figure \ref{fig:Goals_byPlayerType_uncondProb} drives home the apparent outperformance of EPs more clearly. Here we show the share of instances for which we observe EPs and RLPs leaving the pitch with a certain number of goals. Said differently, we plot the unconditional probability of an EP, respectively RLP, to score a certain number of goals in a given match. The unconditional outperformance of EPs can be argued for in two ways -- which are two sides of the same coin: unconditionally, EPs are less likely than RLPs to end a match without having found the back of the net. In reverse, this implies that EPs are much more likely to score at least one goal.

\input{Goals_byPlayerType.tex}

Most of the goals in our sample were actually scored in La Liga, followed by the Premier League and Serie A, as depicted by Figure \ref{fig:Goals_byLeague_bySeason_Total}. However, after normalizing the total number of goals by the number of observations, as done in Figure \ref{fig:Goals_byLeague_bySeason_uncondProb}, it is the Bundesliga in which, on average, an EP scored the most goals. Furthermore, the gap between EPs and RLPs appears to be smallest in Italy's top league -- when, of course, measuring performance only in terms of number of goals. Lastly, Figure \ref{fig:Goals_byLeague_bySeason_Season} shows somewhat hump-shaped distribution in the total number of goals across time, with the distribution for RLPs being slightly left-skewed, suggesting that we draw on slightly more RLP-goals from the earlier part of our sample. Nonetheless, the total number of goals appear to be relatively similarly distributed for EPs and RLPs across seasons, which alleviates any concerns about potential differing inherent structures within our subsample that are not captured by the SFM. 

\input{Goals_byLeague_bySeason.tex}

Similar to the whole ``factor zoo'', the researcher may come up with multiple factors. To showcase the working of the SFM, we keep the number of factors fairly low. This again aligns with the asset-pricing literature in which -- despite the plethora of factors -- the workhorse model is still composed of five\footnote{Often times a sixth factor -- cross-sectional momentum -- \cite{Carhart1997} is added as the theoretical justification for its existence is rather thin, though it has turned out to be a remarkably resilient descriptor of stock returns.} factors only \citep{FF2015}. Still, we achieve a decent performance, and the low dimensional factor setup comes with the benefit of low maintenance on top.

Thus our factor matrix will be composed of only two features. One of those is capturing the \textit{home pitch effect}. This is a binary indicator taking the value 1 if the upcoming match $m$ will be played on the player's home pitch, and 0 otherwise. The other factor is taking into account the differences in points between the player's team and the upcoming opponent, directly proxying for the differences in team strength. We have also experimented with a richer factor structure. However, additional factors such as differences in the goal balance between teams, the rank of the opponent in goals conceded, or the rank of the player's team in goals scored, are already well subsumed by the difference in points among the two teams. As our goal is not just predictive performance, but especially structural interpretability of the model, having a factor structure that is plagued by a high degree of multicollinearity would just lead to unnecessary confusion and result in counterintuitive marginal effects. Nonetheless, if a practitioner can resort to a much richer data set than we have access to, delving into feature engineering and augmenting the set of factors is both advisable and easily implementable in the SFM.

\subsection{SFM Results: Number of Goals Scored per Game}

As already mentioned above, one of the key innovations of the SFM are the evaluation metrics SAR and PAR. Adapted from baseball's WAR framework \citep{BaumerJensenMatthews2015,YurkoVenturaHorowitz2019}, both metrics allow the researcher to compare players along two different dimensions: first, a given player can be compared against a certain \textit{benchmark} -- referred to as the RLP in the WAR literature. Second, players can easily be compared against one-another conditional on being evaluated against the same RLP. Both these analyses can be derived from the panels in Figure \ref{fig:SAR}. 

\vskip 5pt
\noindent{\textsc{\textbf{Skill Above Replacement (SAR).}}} 
More precisely, in Figure \ref{fig:SAR_left} we show SAR for a set of (subjectively picked) EPs. All of the EPs have been evaluated against the RLP which is an artificial player, created as the average of a set of non-EP players, which may nonethless be considered as potential substitutes.  With regards to the two features of SAR outlined above, we see that most of the EPs show higher skill than the RLP -- on average -- as depicted by the white hollow circle.

But thanks to the Bayesian framework, which we embedded the SFM in, we get even more out of SAR than just a simple point estimate.

The handles to the left and right of the mean estimate depict the range within which the outperformance may vary with a probability of 83\%. In essence, the handles document how confident the model is in its point estimate -- and naturally, the model is less weary for players with larger sample size, such as Messi and Cristiano.
Taking a specific example, if we were to only take the point estimate -- the mean -- into consideration, we would infer that Haaland is more skilled than Cristiano Ronaldo. Though, taking the model's\textit{uncertainty} into consideration, we are better advised to not jump to conclusions so quickly.
The handles show that with a probability of 83\%, Haaland's skill allows him to score between $\sim$0.22 and $\sim$0.6 goals per game \textit{more} than the RLP. The lower bound of this range would still allow him to confidently outperform the RLP, and the upper bound would even set him atop of Messi. That's quite a range. In contrast, the model is much more confident in its estimation of Cristiano's skill.
Thus, taking the wide bands around the point estimate for Haaland's $\alpha$ call for caution when settling on a definitive ranking.

Moreover, the handles allow for another interesting point of interpretation, especially when taking a player's age into consideration. For young players, such as Haaland, it shows the up- and downside potential that a team can expect in the future. 

For four players at the bottom of Figure \ref{fig:SAR_left} we still get a mean estimate above zero, but the model assigns an 83\% chance that this outperformance may not hold. In the case of Hazard and Shevchenko, the model sets the players' $\alpha$ below the zero threshold, indicating that the model estimates the players' skill to be lower than the skill of our artificial RLP.

Lastly, forgetting just for a moment the uncertainty around the point estimates, Figure \ref{fig:SAR_left} allows us, here and now, to settle a years', and almost decade's-long, discussion about who is the more skilled: based on SAR alone -- and specifically based on its mean estimate --  Messi does indeed trump Cristiano and takes the title as MSOAT (Most Skilled of All Time).

\input{SAR_v2.tex}

\vskip 5pt
\noindent{\textsc{\textbf{Performance Above Replacement (PAR).}}} 
To recall, SAR quantifies the outperformance of EPs relative to the RLP based on the assumption that there is no heterogeneity in strength across teams. Said differently: all teams are equally strong. Instead, PAR quantifies the same relative outperformance by taking this potential heterogeneity into account. Thus, PAR by itself would just tell us something about the margin to which an EP outperforms the RLP -- including the team differential between the EP's team and the artificial team of the RLP. To what extent such information is useful, is left to the modeler. Yet, a metric that is -- in our point of view -- much more insightful, is the difference between SAR and PAR. This difference, which we show in Figure \ref{fig:SAR_right}, measures the extent to which the team's strength contributes to the player's \textit{observed performance}. The observation $\text{SAR} -\text{PAR} < 0$ would indicate that a player's observed performance is boosted by the strength of the team. If he were to only resort to his skill, he would actually perform worse than what the official statistics tell us. In contrast, $\text{SAR} -\text{PAR} > 0$ suggests that the team is posing a drag on the player's performance, and the player is actually underperforming relative to his ``true'' skill.

Let us again take the example of Haaland: his SAR places him as the second most-skilled player in our select sample. Yet, the SAR-PAR differential suggests that one has to take the SAR value with a grain of salt: his team's strength -- or rather all the teams that he has played for, which in our sample would be Borussia Dortmund and Manchester City --  has been quite a boost to his performance. Hence, the SFM tells us that, yes, Haaland is an exceptionally skilled player, but the teams that he has been playing for, allowed him to perform even better than what he would have been capable of achieving, based on his own skill alone.
Interestingly, the top three players in terms of SAR, are also those that seemed to have benefited the most from the performance of their respective team(s).
This of course, raises the question of whether this is just by chance, or whether there are deeper structural drivers underlying this observation. One simple explanation is the hypothesis that the respective team mates also show superior skill than the ``average'' player, or RLP, and there is a clear amplification mechanism at play: more players with more skill within a team begets better performance of the individual player. An additional explanation could be related to a ``coaching effect''. This would speak to the skill of the \textit{coach}, in that he was capable of understanding which strategy to execute to best utilize and even amplify the player's skill. Since the results in Figure \ref{fig:SAR} are based on a player's whole sample observation, in which he might have had several coaches, such a hypothesis cannot be tested at this stage. Yet, a decomposition by coach and team would be a first step into that direction -- which we leave for future explorations.

As a last point of comparison, let us assess the cross-section in \ref{fig:SAR_right}. For example, Eden Hazard was deemed by the SFM to not really outperform the RLP based on skill alone. However, his observed performance was quite a good representation of his underlying skill, documented by $\text{SAR} \sim\text{PAR}$ in terms of the mean estimate. A similar interpretation holds for Alan Shearer. It is also worth mentioning that taking the uncertainty of the model into consideration, we cannot say with great confidence whether players really benefited from their teams' strength -- in all but two cases: Messi and Cristiano Ronaldo. For those greats, the model is pretty confident that in both cases, the \textit{observed performance} was definitely boosted by the overall strength of the respective team.

Lastly, at this stage, it is worth recalling that in our exercise the SFM carries only two team-factors -- the home-pitch factor, and the point differential between teams. Despite the latter being a reasonable proxy for measuring differences in team strength, additional \textit{uncorrelated} factors would allow for a richer quantification of the team effect. Our current data set, however, does not allow for the addition of other \textit{uncorrelated} factors. Yet, the SFM can be easily augmented with such factors, to which professional sports teams may have access to, based on their proprietary data.

\input{HSGPs_v2.tex}

\vskip 5pt
\noindent{\textsc{\textbf{The Maturity Effect.}}} 
Next we inspect how -- on aggregate -- players' skill evolves over time. In Figure \ref{fig:HSGPs_v2} we show this maturity effect, decomposed into the within-season effect (Figure \ref{fig:HSGPs_v2_within}) and cross-seasonal effect (Figure \ref{fig:HSGPs_v2_across}). The former sheds light on a player's ability to cope with short-term stressors, such as injuries, or exhaustion and fatigue within a single season. The latter shows how, in general, player's skill evolves with the length of a player's career.

In Figure \ref{fig:HSGPs_v2_within}, we thus plot the average of the $h^W\left(D_{i,0:m,s}\right)$ in Equation \eqref{equ:alpha} across all the 144 players in our sample depicted by the yellow solid line. The confidence bands comprise the corresponding 83\% credible interval. Focusing only on the mean estimate, one can see that players perform on average better than expected, at the start of the season. Though after around gameday 15 to 20, this effect starts fade and even turn negative. At the very end of a season, one might anticipate a slight reversal setting in, as players may anticipate the end of the season and tap everything that is left in the tank. Though only marginally visible, the trajectory can be summarized as a ``start-of-season excitement'', a ``mid-season blues'', and an ``end-of-season sprint''. Still, one may argue that a linear approximation of this effect might have been a good fit, as the ``end-of-season sprint'' only imposes slight convexity. Yet, this only holds \textit{on average}. The solid red lines show draws from randomly selected individual players. Here we clearly see the nonlinearities at play. Those can only be captured thanks to the HSGPs.

Figure \ref{fig:HSGPs_v2_across} then shows the longer-term evolution of a player's skill, the $h^C\left(N^S_{i,s}\right)$ parameter in Equation \eqref{equ:alpha}. Here, one can clearly see the nonlinearities at play, as depicted by a humped-shape pattern of the skill parameter over a player's life-cycle. On average, a player's skill seems to reach its peak at around four seasons into the career, before disembarking from its zenith. The delta between the peak at around season 4 and the later seasons is also sizeable, mounting to around 0.25 goals \textit{more per game} that can be expected from a player who is around his zenith relative to a player in his later days.
That this is true on average only, is documented again by the red lines, which are cross-seasonal curves of randomly drawn players. Even though most of them resemble the hump-shaped pattern, some of them show some resurrection towards the end of their career.

Figure \ref{fig:HSGPs_v2_total} , which shows the whole maturity effect as a combination of the within-season and cross-seasonal effects, resembles -- by construction -- the formerly observed dynamics: in general a humped-shaped pattern across seasons, and a steady decline within a single season. This amounts to a player reaching his peak skill -- on average --  after around 150 to 250 games.

\vskip 5pt
\noindent{\textsc{\textbf{Forecasting Ability.}}} 
Lastly, we evaluate the forecasting ability of the SFM. To do so, we compare its predictive power to two simpler -- more reduced-form, i.e. less structural -- versions of itself: (i) \textit{Naive (1)} does keep the SFM's factors, yet the players' $\alpha$ is modeled as just a player-specific intercept term ($\alpha_i$). This specification allows us to understand if accounting for nonlinearities in a player's $\alpha$ also improves its predictive ability; (ii) \textit{Naive (2)} is an even further simplified version of the SFM and only contains a player-specific component, which basically just picks up the unconditional mean of the target, i.e. a player's average \textit{observed performance}. The latter, though appearing to be a very simplistic model, can turn out to be a tough benchmark to beat, especially in an environment where conditional information with predictive value is scarce.

\input{LOO.tex}

For each of those models we run leave-one-out-cross-validation (LOO-CV) -- a method that allows us to exploit the full sample size while ``simulating'' an out-of-sample setting. Yet, it is not a totally hind-sight-bias free environment as the models can leverage information from later observations ($s > t$) to fit the $t^{th}$ observation. Figure \ref{fig:LOO} reports the corresponding \textit{expected log pointwise predictive density} (ELPD) and the corresponding standard deviation of the predictions made on the hold-out sets. The models are ranked from top to bottom as the best to worst performing one. The grey triangle denotes the difference in ELPD relative to the best performing model and the corresponding standard error of the difference.

We see that the SFM is indeed the best performing model. It outperforms the \textit{Naive (1)} benchmark, indicating that not accounting for nonlinearities in the evolution of a player's skill does erode forecasting performance. The SFM also does better than the intercept-only model, \textit{Naive (2)}, suggesting that also the factors do convey important information in how many goals a given player is expected to score in the upcoming match.
Moreover, the superiority of the SFM is also confirmed from a statistical point of view: the difference in the ELPD (hollow circles) between the SFM and the competitors is negative and statistically significant as documented by the grey triangle, with the corresponding standard-errors not crossing the vertical dashed line.

Still, the overall performance is not outstanding. Recall that the SFM can currently only resort to conditioning information of two factors, which surely limits its overall expressive ability. Said differently, the low-dimensional feature vector acts as a strong regularizer, inhibiting the model's flexibility.  

\section{Conclusion} \label{sec:concl}

Inspired by the literature on asset-pricing, we propose the Soccer Factor Model, which is designed to elicit the ``true'' skill of a soccer player -- his $\alpha$. Observed statistics are only a poor proxy for this component, as it is confounded by heterogeneity in strength across teams. 
To this end, the SFM decomposes a player's observed performance into two constituents: team-effort and a player's skill.
Furthermore, we introduce SAR and PAR, two evaluation metrics, which are adaptations of the popular WAR framework (see among others: \cite{BaumerJensenMatthews2015}) that already enjoys widespread application in baseball and American football.

Creating a novel open-source data set, composed of information scraped from the web, we showcase the working of the SFM by predicting the number of goals a striker would score in the upcoming game. 
Mimicking a real-time setting, we fit our model on over 33,000 matches, played by 144 players, in the four major European soccer leagues in season 2000/01 through 2023/24.

Our SAR and PAR metrics allow us to not only settle one of the most fiercely discussed questions in soccer throughout the 2010s -- who is the more skilled? Messi or Cristiano -- but to further understand how different players compare to one another. A fair assessment requires a level-playing-field environment, i.e. a setting unconfounded by any heterogeneity in strength across teams. The SFM allows us to do just that.

Moreover, the Bayesian framework delivers a quantification of uncertainty, i.e. how confident we can be in the estimated point estimate of a player's skill. This uncertainty can further be interpreted as the up- and downside potential -- especially of young and aspiring players.

Lastly, via LOO-CV we show that the SFM does outperform other simpler benchmark models in terms of forecasting performance. 
This lends support to our specific modeling choices, suggesting that (i) accounting for heterogeneity in strength across teams via our \textit{factors}, and (ii) accounting for nonlinear relationships in the evolution of a player's $\alpha$ both have their merit. Not only do they grant the data scientist improved interpretability relative to more reduced-form models, but are also essential for achieving higher forecast accuracy.

All in all, the SFM is an interpretable and adaptable model that allows professional sports teams to better understand and quantify the latent skill of a player -- his $\alpha$.

\clearpage

\bibliographystyle{apalike}
	\bibliography{bib_FA}
	
\clearpage
\appendix

\section*{APPENDIX}

\input{ListOfPlayers.tex}

\end{document}

%% file: Goals_byPlayerType.tex
\begin{figure}[!h]
	\centering
	\caption{\normalsize{Goals by Type of Player}} \label{fig:Goals_byPlayerType}

		\begin{subfigure}[t]{0.5\textwidth}
        \centering   
        	\includegraphics[width=\textwidth, trim = 0mm 5mm 0mm 5mm, clip]{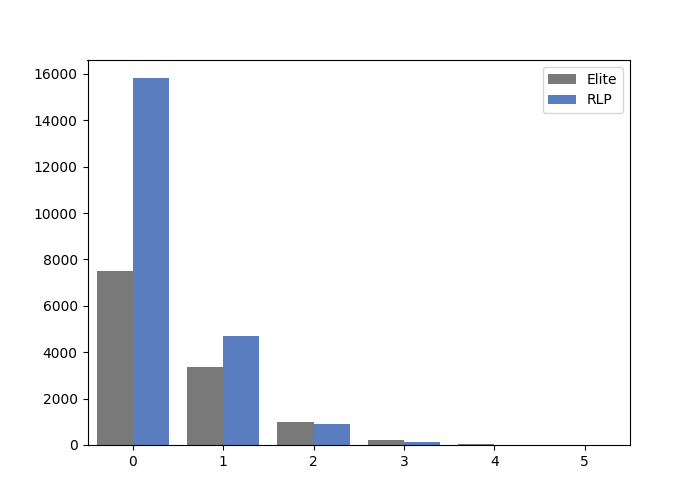}
        \caption{Total Observations by Number of Goals} \label{fig:Goals_byPlayerType_Total}
    \end{subfigure}%
    \begin{subfigure}[t]{0.5\textwidth}
        \centering   
        	\includegraphics[width=\textwidth, trim = 0mm 5mm 0mm 5mm, clip]{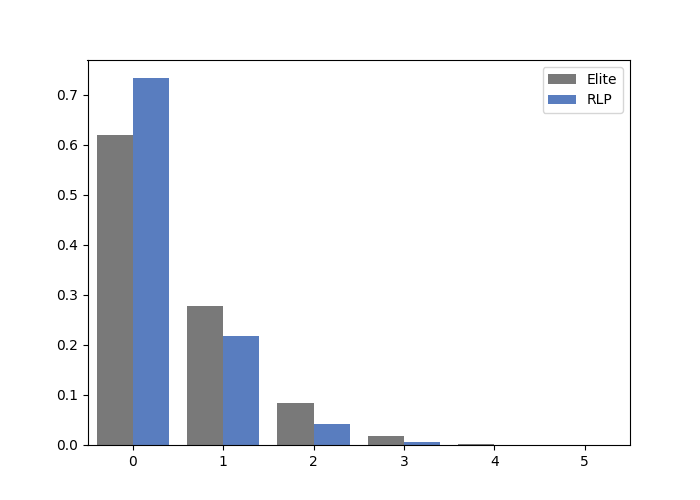}
        \caption{Frequency per Player} \label{fig:Goals_byPlayerType_uncondProb}
    \end{subfigure}

       \begin{threeparttable}
  \begin{tablenotes}
	\scriptsize
		\item[]  \textit{Notes}: On the left hand side, we show the number of instances for which we observe a given number of goals for EPs, and RLPs respectively. On the right hand side, we show the number of instances for which we observe a given number of goals divided by the total number of instances for a specific subgroup of players.
		\end{tablenotes}
\end{threeparttable}
\end{figure}

%% file: Goals_byLeague_bySeason.tex
\begin{figure}[!h]
	\centering
	\caption{\normalsize{Goals by League and Season}} \label{fig:Goals_byLeague_bySeason}

		\begin{subfigure}[t]{0.33\textwidth}
        \centering   
        	\includegraphics[width=\textwidth, trim = 0mm 0mm 0mm 5mm, clip]{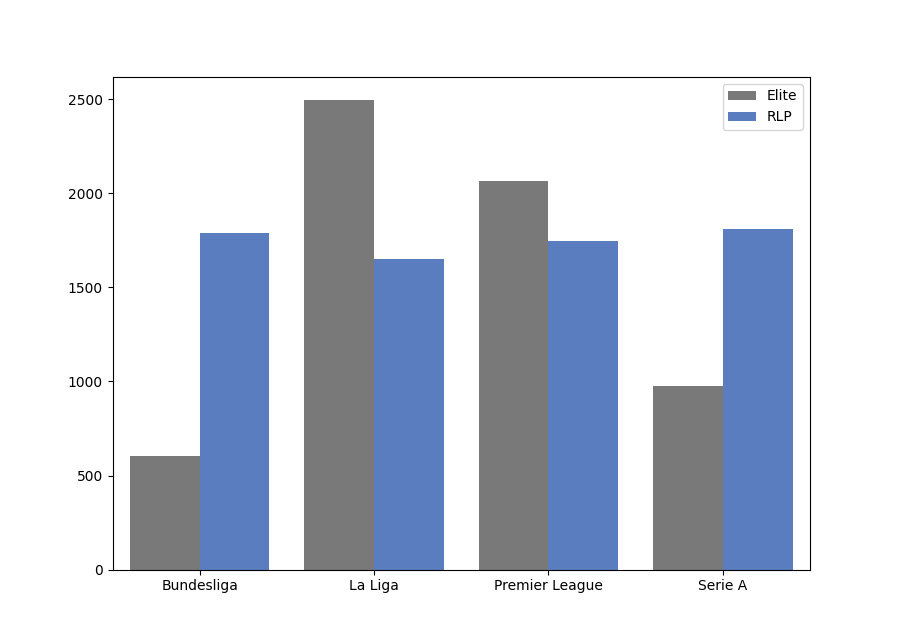}
        \caption{Total Goals} \label{fig:Goals_byLeague_bySeason_Total}
    \end{subfigure}%
    \begin{subfigure}[t]{0.33\textwidth}
        \centering   
        	\includegraphics[width=\textwidth, trim = 0mm 0mm 0mm 5mm, clip]{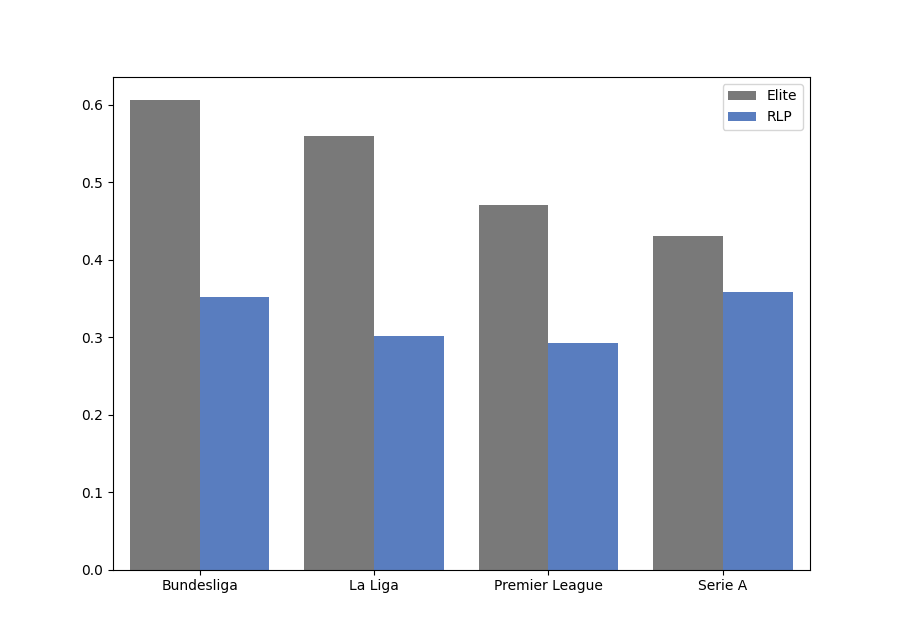}
        \caption{Frequency per Player} \label{fig:Goals_byLeague_bySeason_uncondProb}
    \end{subfigure}
    \begin{subfigure}[t]{0.33\textwidth}
        \centering   
        	\includegraphics[width=\textwidth, trim = 0mm 0mm 0mm 5mm, clip]{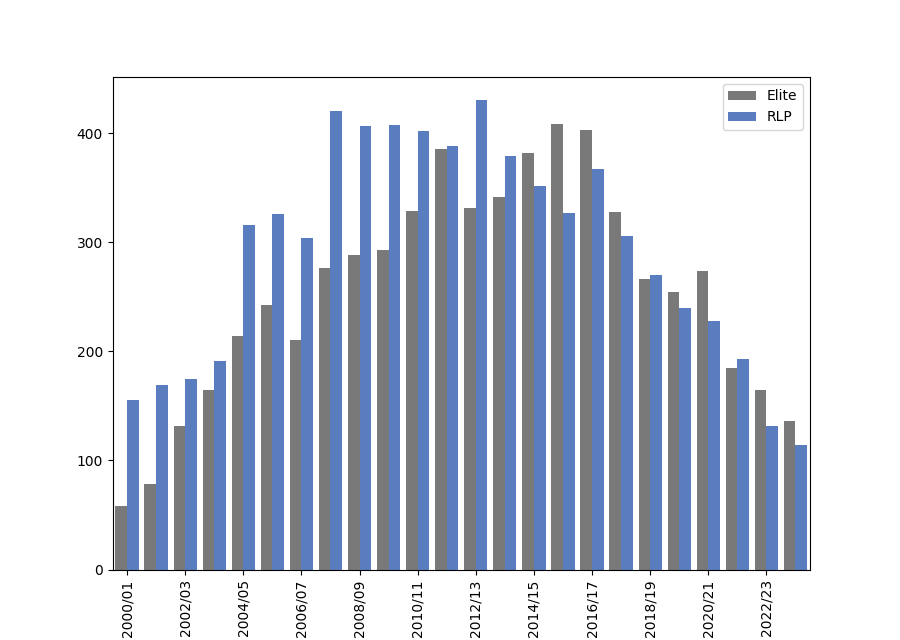}
        \caption{Total Goals by Season} \label{fig:Goals_byLeague_bySeason_Season}
    \end{subfigure}%

       \begin{threeparttable}
  \begin{tablenotes}
	\scriptsize
		\item[]  \textit{Notes}: In panel \ref{fig:Goals_byLeague_bySeason_Total}, we show the total number of goals scored by EPs, respectively RLPs, in any of the four major European leagues. In panel \ref{fig:Goals_byLeague_bySeason_uncondProb}, we divide the total number of goals by the number of sample observations for each subgroup of players and each league. In panel \ref{fig:Goals_byLeague_bySeason_Season} we show the distribution of total goals across time.
		\end{tablenotes}
\end{threeparttable}
\end{figure}

%% file: SAR_v2.tex
\begin{figure}[!h]
	\centering
	\caption{\normalsize{Skill Above Replacement}} \label{fig:SAR}

        \begin{subfigure}[t]{0.5\textwidth}
            \centering   
            \includegraphics[width=0.9\textwidth, trim = 0mm 0mm 0mm 0mm, clip]{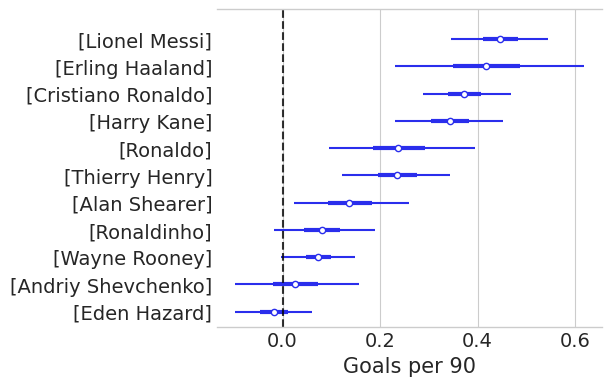}
            \caption{SAR: All Teams Equal}\label{fig:SAR_left}
        \end{subfigure}%
        \begin{subfigure}[t]{0.5\textwidth}
            \centering   
            \includegraphics[width=0.9\textwidth, trim = 0mm 0mm 0mm 0mm, clip]{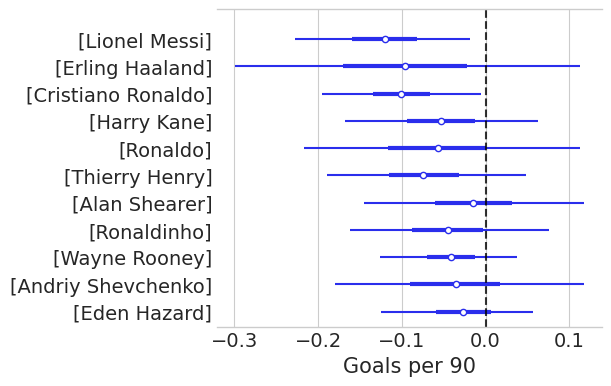}
            \caption{SAR - PAR}\label{fig:SAR_right}
        \end{subfigure}

       \begin{threeparttable}
  \begin{tablenotes}
	\scriptsize
		\item[]  \textit{Notes}: On the left hand side, we show the SAR of EPs in terms of number of goals. That is, the number of goals \textbf{more} per game that an EP would score relative to \textit{the} RLP, only considering a player's $\alpha$, i.e. his skill. On the right hand side we show the difference between SAR and PAR.
		\end{tablenotes}
\end{threeparttable}
\end{figure}

%% file: HSGPs_v2.tex
\begin{figure}[!h]
	\centering
	\caption{\normalsize{Posterior HSGPs}} \label{fig:HSGPs_v2}

        \begin{subfigure}[t]{0.5\textwidth}
            \centering   
            \includegraphics[width=1\textwidth, trim = 0mm 0mm 0mm 0mm, clip]{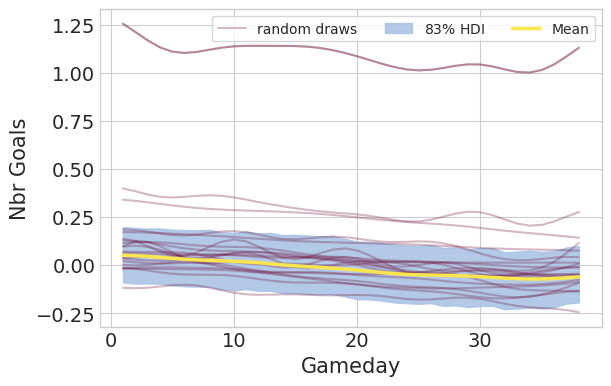}
            \caption{Within-Season Variation}\label{fig:HSGPs_v2_within}
        \end{subfigure}%
        \begin{subfigure}[t]{0.5\textwidth}
            \centering   
            \includegraphics[width=1\textwidth, trim = 0mm 0mm 0mm 0mm, clip]{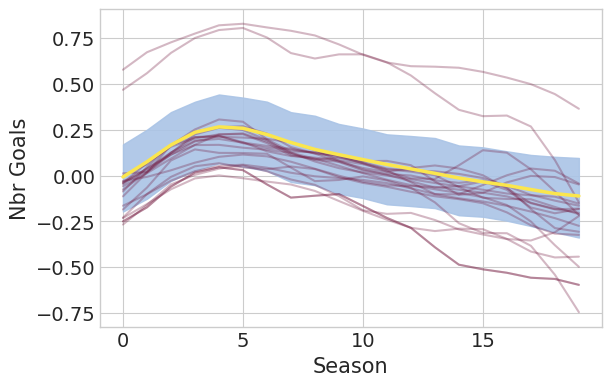}
            \caption{Cross-Seasonal Variation}\label{fig:HSGPs_v2_across}
        \end{subfigure}
        
		\begin{subfigure}[t]{\textwidth}
            \centering   
            \includegraphics[width=\textwidth, trim = 0mm 0mm 0mm 0mm, clip]{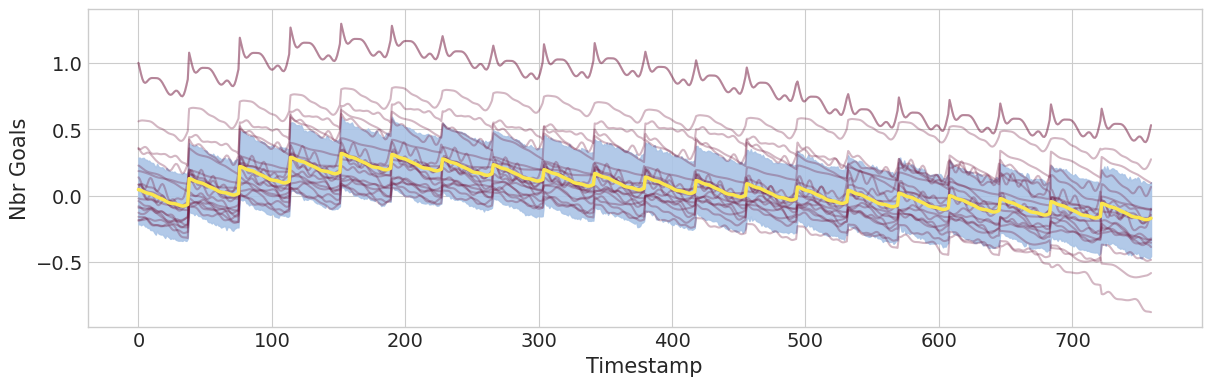}
            \caption{Total GP}\label{fig:HSGPs_v2_total}
        \end{subfigure}

       \begin{threeparttable}
  \begin{tablenotes}
	\scriptsize
		\item[]  \textit{Notes}: In all panels, the solid yellow line depicts the mean across all players and the shaded region encompasses the 83\% credible region around the mean. Panel (a) shows the within-season fluctuations of a player's skill, i.e. how well a player can cope with the progression of a season. Panel (b) shows the evolution of the skill component across seasons, i.e. how the skill of a player evolves over his career. Panel (c) plots the combination of both the within- and cross-seasonal variation as the  sum of the two components. The solid red lines show draws from randomly selected individual players.
		\end{tablenotes}
\end{threeparttable}
\end{figure}

%% file: LOO.tex
\begin{figure}[!h]
	\centering
	\caption{\normalsize{Model Comparison: LOO-CV}} \label{fig:LOO}

    \centering   
     \includegraphics[width=0.9\textwidth, trim = 0mm 0mm 0mm 0mm, clip]{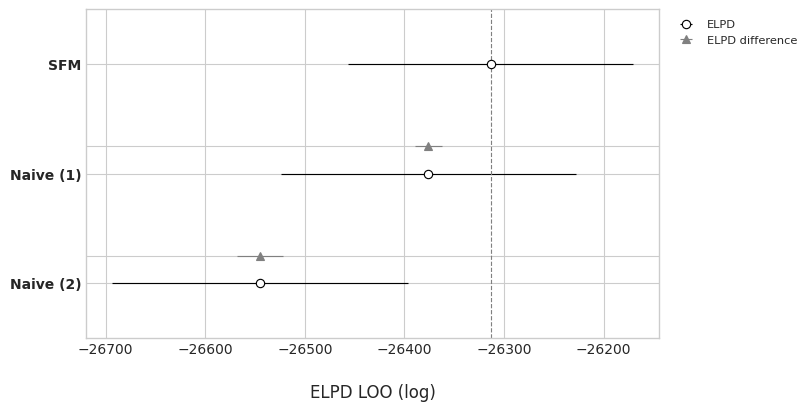}

       \begin{threeparttable}
  \begin{tablenotes}
	\scriptsize
		\item[]  \textit{Notes}: \textit{Expected log pointwise predictive density} (ELPD) for the SFM and two simpler benchmark models. The two benchmarks define $\alpha$ just as a player-specific constant component ($\alpha_i$). Conditioning information of \textit{Naive (1)} is: $\alpha_i$ + \textit{factors}; conditioning information of \textit{Naive (2)} is defined by: $\alpha_i$. 
		\end{tablenotes}
\end{threeparttable}
\end{figure}

%% file: ListOfPlayers.tex
\begin{table}[h!]
\rowcolors{5}{white}{gray!15}
\centering
	\footnotesize
	\centering
	\caption{List of Players} \label{tab:ListOfPlayers}

\resizebox{0.8\textwidth}{!}{%
\begin{tabular}{ll | ll |ll}
\toprule \toprule
\multicolumn{2}{c|}{Elite Players} & \multicolumn{4}{c}{Replacement Level Players} \\
\cmidrule(lr){1-2} \cmidrule(lr){3-6}
Name & Surname  & Name & Surname & Name & Surname \\
\cmidrule(lr){1-2} \cmidrule(lr){3-4} \cmidrule(lr){5-6}
Adriano &  & Choupo-Moting & Eric-Maxim & Kalinic & Nikola \\
Aguero & Sergio & Aduriz & Aritz & Keane & Robbie \\
Bale & Gareth & Ailton &  & Kiessling & Stefan \\
Benzema & Karim & Almeida & Hugo & Klasnic & Ivan \\
Costa & Diego & Altidore & Jozy & Klose & Miroslav \\
Del Piero & Alessandro & Anelka & Nicolas & Lacazette & Alexandre \\
Drogba & Didier & Aspas & Iago & Larsson & Henrik \\
Dybala & Paulo & Aubameyang & Pierre-Emerick & Llorente & Fernando \\
Eto'o & Samuel & Bacca & Carlos & Llorente & Joseba \\
Falcao &  & Barrios & Lucas & Lopez & Adrian \\
Firmino & Roberto & Bellamy & Craig & Makaay & Roy \\
Haaland & Erling & Ben-Yedder & Wissam & Martinez & Jackson \\
Hazard & Eden & Ben-arfa & Hatem & Mertens & Dries \\
Henry & Thierry & Benteke & Christian & Michu &  \\
Hernandez & Javier & Berbatov & Dimitar & Milik & Arkadiusz \\
Higuain & Gonzalo & Berg & Marcus & Milito & Diego \\
Ibrahimovic & Zlatan & Best & Leon & Miller & Kenny \\
Icardi & Mauro & Borriello & Marco & Mintal & Marek \\
Kane & Harry & Briand & Jimmy & Mpenza & Emile \\
Lewandowski & Robert & Cahill & Tim & Neuville & Oliver \\
Lukaku & Romelu & Carew & John & Nolito &  \\
Mandzukic & Mario & Cassano & Antonio & Nugent & David \\
Messi & Lionel & Cavani & Edinson & Olic & Ivica \\
Morata & Alvaro & Crouch & Peter & Paloschi & Alberto \\
Nani &  & Defoe & Jermain & Pazzini & Giampaolo \\
Neymar &  & Del Moral & Manu & Pelle & Graziano \\
Owen & Michael & Dempsey & Clint & Petersen & Nils \\
Pato &  & Di Natale & Antonio & Petric & Mladen \\
Raul &  & Dos Santos & Giovani & Podolski & Lukas \\
Robben & Arjen & Elber & Giovane & Postiga & Helder \\
Robinho &  & Elmander & Johan & Quagliarella & Fabio \\
Ronaldinho &  & Fabiano & Luis & Quaresma & Ricardo \\
Ronaldo & Cristiano & Felix & Joao & Reyes & Jose-Antonio \\
Ronaldo &  & Forlan & Diego & Rosenberg & Markus \\
Rooney & Wayne & F\"ullkrug & Niclas & Sand & Ebbe \\
Salah & Mohamed & Garcia & Sergio & Schieber & Julian \\
Shevchenko & Andriy & Giaccherini & Emanuele & Sergio & Paulo \\
Suarez & Luis & Gilardino & Alberto & Shearer & Alan \\
Torres & Fernando & Giroud & Olivier & Silva & Andre \\
Van Nistelrooy & Ruud & Gomez & Mario & Son & Heung-Min \\
Van Persie & Robin & Gomis & Bafetimbi & Soriano & Jonathan \\
Villa & David & Griezmann & Antoine & Suker & Davor \\
 &  & Guidetti & John & Szalai & Adam \\
 &  & Heskey & Emile & Tomasson & Jon-dahl \\
 &  & Huntelaar & Klaas-Jan & Toni & Luca \\
 &  & Immobile & Ciro & Vardy & Jamie \\
 &  & Insigne & Lorenzo & Vela & Carlos \\
 &  & Inzaghi & Filippo & Vieri & Christian \\
 &  & Jancker & Carsten & Williams & Inaki \\
 &  & Jimenez & Raul & Zamora & Bobby \\
 &  & Joselu &  & Zickler & Alexander \\
\bottomrule \bottomrule
\end{tabular}}

\end{table}

%% file: SFM_v241205.bbl
\begin{thebibliography}{}

\bibitem[Baumer et~al., 2015]{BaumerJensenMatthews2015}
Baumer, B.~S., Jensen, S.~T., and Matthews, G.~J. (2015).
\newblock {openWAR: An Open Source System for Evaluating Overall Player
  Performance in Major League Baseball}.
\newblock {\em Journal of Quantitative Analysis in Sports}, 11(2):69--84.

\bibitem[Berk and van Binsbergen, 2015]{BerkBinsbergen2015}
Berk, J.~B. and van Binsbergen, J.~H. (2015).
\newblock {Measuring Skill in the Mutual Fund Industry}.
\newblock {\em Journal of Financial Economics}, 118(1):1--20.

\bibitem[Carhart, 1997]{Carhart1997}
Carhart, M. (1997).
\newblock {On Persistence in Mutual Fund Performance}.
\newblock {\em The Journal of Finance}, 52(1):57--82.

\bibitem[Coggin et~al., 1993]{CogginFabozziRahman1993}
Coggin, T.~D., Fabozzi, F.~J., and Rahman, S. (1993).
\newblock {The Investment Performance of U.S. Equity Pension Fund Managers: An
  Empirical Investigation}.
\newblock {\em The Journal of Finance}, 48(3):1039--1055.

\bibitem[Diebold et~al., 2023]{DieboldEtAl2023}
Diebold, F.~X., Rudebusch, G.~D., Göbel, M., {Goulet Coulombe}, P., and Zhang,
  B. (2023).
\newblock {When will Arctic Sea Ice Disappear? Projections of Area, Extent,
  Ehickness, and Volume}.
\newblock {\em Journal of Econometrics}, 236(2):105479.

\bibitem[Dixon et~al., 2020]{DixonHalperinBilokon2020}
Dixon, M.~F., Halperin, I., and Bilokon, P. (2020).
\newblock {\em {Machine Learning in Finance: From Theory to Practice}}.
\newblock Springer-Verlags.

\bibitem[Fama and French, 1992]{FF1992}
Fama, E. and French, K. (1992).
\newblock {The Cross-Section of Expected Stock Returns}.
\newblock {\em The Journal of Finance}, 47(2):427--465.

\bibitem[Fama and French, 1993]{FF1993}
Fama, E. and French, K. (1993).
\newblock {Common risk factors in the returns on stocks and bonds}.
\newblock {\em Journal of Financial Economics}, 33(1):3 -- 56.

\bibitem[Fama and French, 2010]{FamaFrench2010}
Fama, E.~F. and French, K.~R. (2010).
\newblock {Luck versus Skill in the Cross-Section of Mutual Fund Returns}.
\newblock {\em The Journal of Finance}, 65(5):1915--1947.

\bibitem[Fama and French, 2015]{FF2015}
Fama, E.~F. and French, K.~R. (2015).
\newblock {A Five-Factor Asset Pricing Model}.
\newblock {\em Journal of Financial Economics}, 116(1):1--22.

\bibitem[Hastie et~al., 2017]{HastieTibshiraniFriedman2017}
Hastie, T., Tibshirani, R., and Friedman, J. (2017).
\newblock {\em {The Elements of Statistical Learning (2nd Edition)}}.
\newblock Springer-Verlag.

\bibitem[Micchelli et~al., 2006]{MicchelliXuZhang2006}
Micchelli, C.~A., Xu, Y., and Zhang, H. (2006).
\newblock {Universal Kernels}.
\newblock {\em Journal of Machine Learning Research}, 7(95):2651--2667.

\bibitem[Murphy, 2023]{Murphy2023}
Murphy, K.~P. (2023).
\newblock {\em {Probabilistic Machine Learning: Advanced Topics}}.
\newblock The MIT Press.

\bibitem[Rasmussen and Williams, 2006]{RasmussenWilliams2006}
Rasmussen, C. and Williams, C. (2006).
\newblock {\em {Gaussian Processes for Machine Learning}}.
\newblock The MIT Press.

\bibitem[Riutort-Mayol et~al., 2022]{RiutortMayolEtAl2022}
Riutort-Mayol, G., B{\"u}rkner, P.-C., Andersen, M.~R., Solin, A., and Vehtari,
  A. (2022).
\newblock {Practical Hilbert space approximate Bayesian Gaussian processes for
  probabilistic programming}.
\newblock {\em Statistics and Computing}, 33(1).

\bibitem[Yurko et~al., 2019]{YurkoVenturaHorowitz2019}
Yurko, R., Ventura, S., and Horowitz, M. (2019).
\newblock {nflWAR: A Reproducible Method for Offensive Player Evaluation in
  Football}.
\newblock {\em Journal of Quantitative Analysis in Sports}, 15(3):163--183.

\end{thebibliography}
